# Non-Abelian Generalizations of the Hofstadter model: Spin-orbit-coupled Butterfly Pairs

Yi Yang[1],* Bo Zhen[2], John D. Joannopoulos[1], and Marin Soljačić[1]

[1] Department of Physics and Research Laboratory of Electronics, Massachusetts Institute of Technology, Cambridge, Massachusetts 02139, USA

[2] Department of Physics and Astronomy, University of Pennsylvania, Philadelphia, Pennsylvania 19104, USA

## Abstract

The Hofstadter model, well known for its fractal butterfly spectrum, describes two-dimensional electrons under a perpendicular magnetic field, which gives rise to the integer quantum Hall effect. Inspired by the real-space building blocks of non-Abelian gauge fields from a recent experiment, we introduce and theoretically study two non-Abelian generalizations of the Hofstadter model. Each model describes two pairs of Hofstadter butterflies that are spin-orbit coupled. In contrast to the original Hofstadter model that can be equivalently studied in the Landau and symmetric gauges, the corresponding non-Abelian generalizations exhibit distinct spectra due to the non-commutativity of the gauge fields. We derive the genuine (necessary and sufficient) non-Abelian condition for the two models from the commutativity of their arbitrary loop operators. At zero energy, the models are gapless and host Weyl and Dirac points protected by internal and crystalline symmetries. Double (8-fold), triple (12-fold), and quadrupole (16-fold) Dirac points also emerge, especially under equal hopping phases of the non-Abelian potentials. At other fillings, the gapped phases of the models give rise to topological insulators. We conclude by discussing possible schemes for experimental realization of the models on photonic platforms.

## Introduction

Synthetic gauge fields[1] provide a versatile toolbox to manipulate geometric phases in engineered physical systems. These gauge fields can be classified into Abelian (commutative) and non-Abelian (non-commutative) depending on the commutativity of the underlying group. Much success has been achieved in synthesizing Abelian gauge fields on different platforms, including cold atoms[2-10], photons[11-22], phonons[23-25], polaritons[26], and superconducting qubits[27-29]. It is more demanding to synthesize non-Abelian gauge fields because they require internal degrees of freedom and non-commutative matrix-valued gauge potentials. Nevertheless, remarkable progress has recently been made. In momentum space, non-Abelian gauge fields manifest as high-dimensional spin-orbit coupling[30, 31] in atoms[32, 33], liquid crystals[34], and exciton-polaritons[35-36]. They also affect the wavepacket evolution in the momentum space of mechanical lattices[37]. In addition, non-Abelian topology can also manifest in interconnected nodal lines[38, 39] and multiband charges[38, 40]. In synthetic spaces, the spin evolution of nuclear magnetic resonances[41-44] is an early example exhibiting non-Abelian geometric phases[45, 46]. Such phases have been utilized for implementing non-Abelian holonomic quantum gates[47] and non-Abelian Yang monopoles[48-49]. In real space, the non-Abelian Aharonov–Bohm effect[50-52] has recently been observed in a minimal-scheme interference measurement[53], where temporal modulation and the Faraday effect provide artificial magnetic fields for the pseudospin of light of different orientations. It is natural to ask what physics could be enabled by the recently demonstrated real-space non-Abelian gauge fields in larger systems such as lattices.

*Correspondence: Yi Yang (yiy@mit.edu)

A widely celebrated lattice model, featuring gauge fields, is the Hofstadter butterfly[54]. This model depicts particles on a two-dimensional square lattice threaded by a uniform magnetic field, which is described by a U(1) Abelian gauge field. This Abelian model has been widely studied theoretically and realized experimentally with atoms[4, 6], photons[55], and superconducting qubits[56] in real space. More recently, edge features of this model have also been implemented in a synthetic frequency ladder of an optical ring resonator[57]. To generalize the Hofstadter model to non-Abelian versions, one needs to substitute its scalar hopping phase with matrix-valued gauge potentials. Studies along this direction have yielded interesting phenomena, such as the non-Abelian Hofstadter moth spectrum[51], the integer quantum Hall effect under constant lattice Wilson loops[58], the quantum spin Hall effect[59], and the associated non-Hermitian generalizations[60].

Inspired by the recent experiment[53] demonstrating building blocks of real-space non-Abelian gauge fields $\theta\sigma_y$ and $\phi\sigma_z$, here, we introduce and systematically study two distinct non-Abelian Hofstadter models in the Landau [$H^L$, Eq. (4)] and symmetric [$H^s$, Eq. (6)] gauges. Different from previous non-Abelian generalizations[51, 58-62], our models describe two pairs of Hofstadter butterflies that are spin-orbit coupled. We analytically derive the genuine (necessary and sufficient) non-Abelian condition shared by our two models by examining the commutativity of arbitrary loop operators. The spectra of our models reduce to multiple copies of the Hofstadter butterfly when the Abelian condition is met; otherwise, they change substantially—they inherit the fractal nature but exhibit modified butterfly features. When chiral symmetry is present, our non-Abelian models host Dirac and Weyl points at zero energy that are stabilized by internal and spatial symmetries. When the non-Abelian hopping phases become equal (i.e., $\theta = \phi$), high-degeneracy points appear, such as double (8-fold), triple (12-fold), and quadruple (16-fold) Dirac points. Additionally, we observe a dependence between the appearance of Dirac points at the time-reversal-invariant momenta (TRIMs) and the hopping phases of the non-Abelian gauge fields [see Eq. (16)]. At other fillings, the bulk gaps of the models exhibit topological insulating phases with helical edge states. Finally, we discuss possible experimental realizations of the models on photonic platforms and their generalization to higher dimensions.

## Results

### Models

We start by reviewing the Abelian Hofstadter model[54], as illustrated in Fig. 1a. This model describes two-dimensional spinless particles in a uniform perpendicular magnetic field. Therefore, it is a hallmark of quantum Hall physics. Its bulk gaps possess non-zero Chern numbers and host chiral edge states. Its Abelian version can be described by the Hofstadter–Harper Hamiltonian

$$H^0(\phi) = -\sum_{m,n} t_x c^\dagger_{m+1,n} c_{m,n} + t_y c^\dagger_{m,n+1} e^{im\phi} c_{m,n} + \text{H.c.} \quad (1)$$

where its Abelian gauge potential reads

$$\mathbf{A} = (0, m\phi, 0) \quad (2)$$

Here, $t_{x,y}$ are the real hopping terms in the $x$ and $y$ directions, and we restrict ourselves to the case of $t_x = t_y = t$ throughout the paper (including for the non-Abelian models below). $c_{m,n}$ and $c^\dagger_{m,n}$ are the annihilation and creation operators of site $(m,n)$. $\phi = 2\pi\varphi/\varphi_0$ is the Peierls hopping phase, where $\varphi = \iint \mathbf{B} \cdot \mathrm{d}\mathbf{S}$ is the magnetic flux per unit cell and $\varphi_0 = h/e$ is the flux quanta. When $\phi/2\pi$ is a rational number, i.e., $\phi = 2\pi p_\phi/q_\phi$, where $p_\phi$ and $q_\phi$ are integers, the system is translationally invariant in a $q_\phi \times 1$ magnetic unit cell, defined as $q_\phi$ units along $x$ and 1 unit along $y$. Accordingly, the associated band structure is defined in the magnetic Brillouin zone (MBZ) where $k_x \in [0, 2\pi/q_\phi)$ and $k_y \in [0, 2\pi)$. The eigenspectrum of this Hamiltonian is the Hofstadter butterfly, a famous example of quantum fractals.

The original Abelian Hamiltonian [Eq. (1)] is written in the Landau gauge. Still within this Landau gauge, we revise the U(1) gauge potential Eq. (2) into an SU(2) gauge potential that reads

$$\mathbf{A} = (0, m\theta\sigma_y + m\phi\sigma_z, 0) \quad (3)$$

where $\sigma_y$ and $\sigma_z$ are Pauli matrices and $\theta, \phi \in [0, 2\pi)$ are again hopping phases. The associated Hamiltonian for spinful particles is thus given by

$$H^L(\theta, \phi) = -\sum_{m,n} t_x c^\dagger_{m+1,n} c_{m,n} + t_y c^\dagger_{m,n+1} e^{im\theta\sigma_y} e^{im\phi\sigma_z} c_{m,n} + \text{H.c.}$$

$$= -\sum_{m,n} t_x (c^\dagger_{\uparrow,m+1,n}, c^\dagger_{\downarrow,m+1,n}) \begin{pmatrix} c_{\uparrow,m,n} \\ c_{\downarrow,m,n} \end{pmatrix} \quad (4)$$

$$+ t_y (c^\dagger_{\uparrow,m,n+1}, c^\dagger_{\downarrow,m,n+1}) \begin{pmatrix} \cos m\theta & \sin m\theta \\ -\sin m\theta & \cos m\theta \end{pmatrix} \begin{pmatrix} \exp im\phi & 0 \\ 0 & \exp -im\phi \end{pmatrix} \begin{pmatrix} c_{\uparrow,m,n} \\ c_{\downarrow,m,n} \end{pmatrix} + \text{H.c.}$$

where we show both spin-index implicit and explicit expressions. The schematic of this model is shown in Fig. 1b, which we refer to as the Landau-gauge model hereafter. If $\theta/2\pi$ and $\phi/2\pi$ are both rational, i.e., they can be written as $\theta = 2\pi p_\theta/q_\theta$ and $\phi = 2\pi p_\phi/q_\phi$, where $p_\theta, q_\theta, p_\phi$, and $q_\phi$ are integers, then Eq. (4) can be solved numerically in the MBZ $k_x \in [0, 2\pi/q)$ and $k_y \in [0, 2\pi)$, where $q = \text{lcm}(q_\theta, q_\phi)$ is the least common multiple of $q_\theta$ and $q_\phi$.

Alternatively, we can adopt the symmetric gauge and set the non-Abelian phases in different directions, as shown in Fig. 1c. The associated gauge potential reads

$$\boldsymbol{A} = (-n\theta\sigma_y, m\phi\sigma_z, 0) \tag{5}$$

The associated Hamiltonian is given by

$$H^s(\theta, \phi) = -\sum_{m,n} t_x c^\dagger_{m+1,n} e^{-in\theta\sigma_y} c_{m,n} + t_y c^\dagger_{m,n+1} e^{im\phi\sigma_z} c_{m,n} + \text{H.c.}$$

$$= -\sum_{m,n} t_x \left(c^\dagger_{\uparrow,m+1,n}, c^\dagger_{\downarrow,m+1,n}\right) \begin{pmatrix} \cos n\theta & -\sin n\theta \\ \sin n\theta & \cos n\theta \end{pmatrix} \begin{pmatrix} c_{\uparrow,m,n} \\ c_{\downarrow,m,n} \end{pmatrix} \tag{6}$$

$$+ t_y \left(c^\dagger_{\uparrow,m,n+1}, c^\dagger_{\downarrow,m,n+1}\right) \begin{pmatrix} \exp im\phi & 0 \\ 0 & \exp -im\phi \end{pmatrix} \begin{pmatrix} c_{\uparrow,m,n} \\ c_{\downarrow,m,n} \end{pmatrix} + \text{H.c.}$$

where spin-index implicit and explicit expressions are also shown. We refer to this model as the symmetric-gauge model hereafter. Still adopting the rationality assumption, the Hamiltonian can be solved in the $q_\phi \times q_\theta$ supercell with the MBZ defined by $k_x \in [0, 2\pi/q_\phi)$ and $k_y \in [0, 2\pi/q_\theta)$.

We emphasize that although we label the two models as 'Landau-gauge' and 'symmetric-gauge' according to the arrangement of their link variables, the two models $H^L$ and $H^s$ are physically distinct because of the non-commutativity of the gauge potentials. This is different from the Abelian Hofstadter model $H^0$ that is equivalent in the Landau and symmetric gauges. In addition, for either $H^L$ or $H^s$, its spectra and the associated topological phenomena still hold if $\theta$ and $\phi$ are interchanged.

**Genuine non-Abelian conditions**

In this section, we derive the genuine non-Abelian conditions for the two models. Furthermore, we show that if the models reduce to Abelian, then equivalently, their spectra also reduce to the original Hofstadter butterfly.

Precisely defining non-Abelian gauge fields can be complex, as several criteria exist in the literature, which are not equivalent to each other. For example, the non-commutativity of the gauge potentials $[A_\mu, A_\nu] \neq 0$, or that of their link variables $[U_\mu, U_\nu] \neq 0$, where $U_\mu \equiv e^{iA_\mu}$, is sometimes referred to as the criterion for non-Abelian gauge fields[63, 64]. A different criterion concerns the gauge-invariant Wilson loop, i.e., the trace of the loop operator $\boldsymbol{W}$. Such a loop operator $\boldsymbol{W}$ is defined as $\boldsymbol{W} = \exp(i\oint \boldsymbol{A} \cdot d\boldsymbol{l})$. In a square lattice, the loop operator $\boldsymbol{W}(\boldsymbol{r})$ for a unit plaquette, with site $\boldsymbol{r} = (m, n)$ at the bottom left corner, is explicitly given by

$$W(r) = U_y^\dagger(r)U_x^\dagger(r+\hat{e}_y)U_y(r+\hat{e}_x)U_x(r) \tag{7}$$

where $\hat{e}_\mu$ is the unit vector in the $\mu$ direction, and we adopt the counterclockwise convention. The phase of $W(r)$ is the real-space non-Abelian Berry curvature (magnetic field) $F_{xy}(r)$. While $W(r)$ and $F_{xy}(r)$ are generally not gauge invariant, the Wilson loop $W(r) \equiv \text{Tr}\, W(r)$ is. Accordingly, non-Abelian gauge fields can be defined as $|W| \neq N$ for $N$-fold degenerate systems. Nevertheless, the aforementioned three criteria, namely, $[A_\mu, A_\nu] \neq 0$, $[U_\mu, U_\nu] \neq 0$, and $|W| \neq N$, are all necessary but insufficient conditions for non-Abelian gauge fields[64].

It has been established that the necessary and sufficient condition for non-Abelian gauge fields is that two spatially connected loop operators $W_1$ and $W_2$ are not commutative (alternatively, non-commutative Berry curvatures $[F_{\mu\nu}(r), F_{\mu\nu}(r)] \neq 0$)[64]. Based on this, we examine the commutativity of arbitrary loop operators of unit plaquettes to derive the genuine non-Abelian conditions for our lattice models (see Materials and methods A).

For compact notation, we define $\Theta_m \equiv \exp(im\theta\sigma_y)$ and $\Phi_m \equiv \exp(im\phi\sigma_z)$ for the link variables. We also set $\Theta \equiv \Theta_1$ and $\Phi \equiv \Phi_1$. For our non-Abelian models $H^L$ and $H^S$, we rigorously prove that they share the same necessary and sufficient non-Abelian condition

$$[\Theta\Phi, \Phi\Theta] \neq 0 \tag{8}$$

The detailed proof is contained in Materials and methods A. Below, we only summarize the key steps.

The link variables at site $(m,n)$ of $H^L$ in the $y$ direction are given by $L_{m,n}^L = \Theta_m\Phi_m$. Lemma A.1 proves that $L_{m,n}^L$ orms an Abelian group if and only if $[\Theta\Phi, \Phi\Theta] = 0$. The loop operators for a unit plaquette [bottom left corner at coordinates $(m,n)$] in $H^L$ and $H^S$ can be obtained as

$$W_{m,n}^L = \Phi_{-m}\Theta\Phi_{m+1} \tag{9}$$

$$W_{m,n}^S = \Phi_{-m}\Theta_{n+1}\Phi_{m+1}\Theta_{-n} \tag{10}$$

We prove that $W_{m,n}^L$ and $W_{m,n}^S$ both become Abelian groups if and only if $L_{m,n}^L$ is an Abelian group, which is equivalent to $[\Theta\Phi, \Phi\Theta] = 0$ (shown in Lemma A.1 in Materials and methods A).

Simple algebra yields

$$[\Theta\Phi, \Phi\Theta] = \begin{pmatrix} 2i\sin^2\theta\sin2\phi & -2\sin2\theta\sin^2\phi \\ 2\sin2\theta\sin^2\phi & -2i\sin^2\theta\sin2\phi \end{pmatrix} \tag{11}$$

We can therefore find that the conditions for the systems to reduce to Abelian are $\theta \in \{0, \pi\}$, $\phi \in \{0, \pi\}$, or $\{\theta, \phi\} \in \{\pi/2, 3\pi/2\}$. For comparison, we recall that in a recent non-Abelian Aharonov--Bohm experiment[53], the Abelian condition is that either $\theta$ or $\phi$ is a multiple integer of $\pi/2$. The difference between the two conditions arises from the fact that Ref. 53 only examines the commutativity between time-reversal pairs of the loop operators, while here, the lattice models involve many more loop operators of all the plaquettes. As a result, a larger part of the parameter space $(\theta, \phi)$ falls into the non-Abelian regime for the lattice models.

Being Abelian is also the necessary and sufficient condition for two decoupled Hofstadter butterflies to emerge in the eigenspectra (see the bulk spectra in the supplementary videos). We first consider $H^L$, which has matrix elements given by $H^L_{m,m} = e^{ik_y}\Theta_m\Phi_m + \text{H.c.}$ and $H^L_{m,m-1} = e^{ik_x}\sigma_0$, where $\sigma_0$ is the identity matrix. We want to diagonalize $H^L$ with a set of local gauge transformations $U = \text{diag}[U_1, U_2, ..., U_q]$ such that $UH^LU^\dagger = H^L(\theta = 0, \phi)$, which obviously enables a pair of decoupled butterflies to be generated by opposite magnetic fields. To diagonalize $H^L_{m,m}$, $U_m$ should be eigenvectors of $H^L_{m,m}$. On the other hand, the off-diagonal term $H^L_{m,m-1}$ needs to remain invariant, which requires $U_1 U_2^\dagger = \cdots = U_{q-1}U_q^\dagger = U_q U_1^\dagger = \sigma_0$. Thus, we have $U_1 = U_2 = \cdots = U_q$. Therefore, to have a pair of decoupled butterflies in the spectrum of $H^L$, there should exist a global unitary transformation $U$ that simultaneously diagonalizes all block diagonal elements $H^L_{m,m}$. This is equivalent to the requirement that all $H^L_{m,m}$ commute, i.e., $[e^{ik_y}\Theta_m\Phi_m + \text{H.c.}, e^{ik_y}\Theta_{m'}\Phi_{m'} + \text{H.c.}] = 0$ for an arbitrary choice of m, m', and $k_y$. In Materials and methods B, we further show that this requirement is equivalent to the genuine Abelian condition $[\Theta\Phi, \Phi\Theta] = 0$.

For the symmetric-gauge model $H^s$, a local gauge transformation $U_{m,n} = \Theta_{-mn}$ can be applied, which eliminates the link variables in the $x$ direction and transforms the link variables in the $y$ direction between sites $(m, n)$ and $(m, n + 1)$ as $L^s_{m,n} = \Theta_{-mn}\Phi_m\Theta_{m(n+1)}$. For the transformed Hamiltonian, the argument for $H^L$ in the previous paragraph can similarly prove that the spectral re-emergence of the Hofstadter butterfly is equivalent to the condition that $L^s_{m,n}$ forms an Abelian group. In Materials and methods B, we prove that this requirement is also equivalent to the genuine Abelian condition $[\Theta\Phi, \Phi\Theta] = 0$.

**Bulk spectra, gapless zero modes, and gapped phases**

Since both models $H^L$ and $H^s$ are invariant under the interchange of $\theta$ and $\phi$, we show the bulk spectra $E(\theta, \phi)$ of the two models for different surface cuts along the $\theta$ direction in Fig. 2 (also see the bulk spectra in the supplementary videos). For $\theta/2\pi \in \{0, 1/2\}$ (Fig. 2a, d, e, and h), both $H^L$ and $H^s$ are Abelian, and the spectra restore the Hofstadter butterfly with two-fold degeneracy. For $\theta = \pi$, the butterfly simply translates by $\pi$ in the $\phi$ axis direction. For $\theta/2\pi \in \{1/4, 1/3\}$ (Fig. 2b, c, f, and g), both $H^L$ and $H^s$ are non-Abelian but exhibit distinct spectra. Similar to the Abelian case, repetitions of similar structures at various scales still appear. A variety of 'butterfly-like' structures exhibit either large gaps or small gaps.

We now address the internal symmetries of the non-Abelian models. Different from the Abelian Hofstadter model $H^0$, which breaks the spinless time-reversal symmetry $K$, the two Hamiltonians $H^L$ and $H^s$ both obey the fermionic spinful time-reversal symmetry $i\sigma_y K$. In momentum space, the Abelian model $H^0$ has a 'translational' symmetry $E(k_x, k_y) \to -E(k_x + \pi, k_y + \pi)$[7], which is inherited by $H^L$ and $H^s$ regardless of the choice of $(\theta, \phi)$. We show this in Materials and methods C from the associated Harper equations of the two non-Abelian models. For chiral symmetry, we recall that the Abelian Hofstadter model $H^0$ obeys chiral symmetry when $q_\phi$ is even[65]. The chiral operator $S^0$ for $H^0$ is given explicitly by $(S^0 \Psi)_m = (-1)^m (\mathrm{i})^{q/2} \Psi_{m+q/2}$[64], where $\Psi_m$ is the $m$th component of the wavefunction. In the non-Abelian models $H^L$ and $H^s$, the condition for chiral symmetry is modified. Specifically, $H^L$ obeys chiral symmetry when $q_\theta q_\phi$ is even and $p_\theta \tilde{q}_\phi + p_\phi \tilde{q}_\theta$ is odd, where $\tilde{q}_{\theta(\phi)} \equiv q_{\theta(\phi)}/\gcd(q_\theta, q_\phi)$ and gcd is the greatest common divisor (see Materials and methods C); $H^s$ obeys chiral symmetry when $q_\theta q_\phi$ is even (see Materials and methods C). Therefore, the condition for chiral symmetry is relatively weak for the symmetric-gauge model $H^s$. The original chiral symmetry operator $S^0$ is also inherited by $H^L$ and $H^s$ with slight modifications. The explicit, modified chiral operators $S^L$ and $S^s$ are given by

$$(S^L \Psi)_m = (-1)^m (\mathrm{i})^{q/2} \sigma_0 \Psi_{m+q/2} \tag{12}$$

$$(S^s \Psi)_{m,n} = (-1)^m (\mathrm{i})^{q_\phi/2} \sigma_0 \Psi_{m+q_\phi/2, n} \tag{13}$$

In Eq. (13), we assume that $q_\phi$ is even without loss of generality. This chiral symmetry plays a pivotal role in the gapless zero modes of the models. Throughout the paper, we refer to two- and four-fold linear band crossings as Weyl and Dirac points, respectively, although our models are two-dimensional.

We review the Weyl points in the Abelian Hofstadter model $H^0$. It is established that when $q_\phi$ is even, $H^0$ has chiral symmetry, which results in $q_\phi$ numbers of Weyl points at zero energy (Fig. 3a and b). In a basis where the chiral operator is diagonal, the Hamiltonian is block off-diagonal, $H = (0, h; h^\dagger, 0)$. The determinant $D \equiv \det h$ of the 'reduced' Hamiltonian $h$ enables the definition of a $\mathbb{Z}_2$ winding number

$$\nu_k \equiv \frac{1}{2\pi \mathrm{i}} \oint_C \frac{1}{D} \mathrm{d}D \bmod 2 \tag{14}$$

where $C$ is an infinitesimal loop around momentum $\boldsymbol{k}$. $\nu_k$ dictates the Weyl point momenta. Specifically, Weyl points emerge at k where $\nu_k = 1$, i.e., Re $D = 0$ and Im $D = 0$ are simultaneously satisfied[65]. For the Abelian Hofstadter model, $H^0$, Re $D^0$ and Im $D^0$ are simple sine or cosine functions of $k_x$ or $k_y$, as shown in Fig. 3a2, a3, b2, and b3. Moreover, $\nu_{\pi/2, \pi/2} = 1$ is guaranteed, which leads to $q_\phi$ numbers of Weyl points in total, including the translational invariance. Two scenarios can be classified by the winding number at the Γ point. In particular, for $q_\phi = 4\mathbb{Z}$, $\nu_\Gamma = 1$, and the Weyl points are located at the edges of the MBZ (Fig. 3a2 and a3). For $q_\phi = 4\mathbb{Z} + 2$, $\nu_\Gamma = 0$, and the Weyl points are located inside the MBZ (Fig. 3b2 and b3). These degeneracies have a $\mathbb{Z}$ classification, as $H^0$ belongs to the symmetric class AIII[66].

The above analysis, based on the winding number, can be similarly applied to the non-Abelian models $H^L$ and $H^s$, as shown in Fig. 3c-f. In the presence of spin-orbit coupling, the two models are both gapless at zero energy and host even numbers of Weyl and/or Dirac points when chiral symmetry is present. For $H^L$, aside from the Dirac points at TRIMs, two pairs of Weyl points (solid circles in Fig. 3c2 and c3) appear at non-TRIM points inside the MBZ. These Weyl nodes appear because of the presence of time-reversal symmetry and the lack of inversion symmetry in $H^L$, which, taken together, requires $4\mathbb{Z}$ numbers of Weyl degeneracies[67] for a zero Chern number in total. This simultaneous appearance of Dirac and Weyl points provides a simple tight-binding realization of a Dirac-Weyl semimetal[68] in two dimensions. On the other hand, $H^s$ has both time-reversal and inversion symmetries and is therefore Kramer partnered over the entire MBZ. As a result, all the linear nodes are Dirac points, whether at the edge (Fig. 3e) or inside (Fig. 3f) the MBZ.

$H^L(2\pi/3, \pi/2)$ (Fig. 3c) belongs to class CII. Its Weyl nodes, located at non-TRIM points (solid circles in Fig. 3c3), have a $2\mathbb{Z}$ classification. Its Dirac points, located at TRIM points (open circles in Fig. 3c3), have a $\mathbb{Z}_2$ classification. These Weyl nodes are locally stable against perturbations that preserve the internal symmetries, while the Dirac points are not. For example, an on-site potential perturbation $\Delta\lambda_m^L = (-1)^{m<=q/2}\Delta$ [where $q = \mathrm{lcm}(q_\theta, q_\phi)$], which respects all internal symmetries, splits the two Dirac points in Fig. 3c into four Weyl nodes towards non-TRIM points. The original 4 Weyl nodes in Fig. 3c are locally robust against such a perturbation. Indeed, the Dirac points of $H^L$ are stabilized by inversion symmetry only at TRIMs. This inversion operator commutes with both $H^L$ and the chiral operator $S^L$. Therefore, both $H^L$ and $S^L$ can be simultaneously block diagonalized and labelled by the inversion eigenvalues[69]. The degeneracy of such chiral zero modes at an inversion-invariant momentum k is at least

$$N_\mathbf{k}^L(\theta, \phi) = \left|\mathrm{tr}S_\mathbf{k}^L(+; \theta, \phi)\right| + \left|\mathrm{tr}S_\mathbf{k}^L(-; \theta, \phi)\right| \tag{15}$$

where the superscript labels the Hamiltonian ($H^L$ or $H^s$) and $\pm$ labels the inversion eigenvalues. We find that $N_{\Gamma,Y}^L(2\pi/3, \pi/2) = 4$, which corresponds to the Dirac points in Fig. 3c. We verify the protection with a site-dependent perturbation $\Delta t_m^L = (-1)^m \Delta \sigma_z$ on the tunneling amplitude $t$, which breaks the time-reversal $T$, particle-hole $P$, and reflection $M_x$ symmetries while respecting the chiral $S$ and inversion $I$ symmetries at TRIMs. Under this perturbation, the Dirac points of $H^L$ remain pinned at TRIMs, indicating their protection by chiral and inversion symmetries.

Different from $H^L$, $H^s$ has inversion symmetry. The inversion operator $I$ of $H^s$ commutes with both the time-reversal $T$ and particle-hole $P$ symmetries. It also has reflection symmetry $M_x$ (chosen to be $\sigma_x$ to meet the Hermitian requirement[70]) that anticommutes with $T$ and $P$. When $q_\theta$ and $q_\phi$ are both even [e.g., $(\theta, \phi)/2\pi = (1/4, 1/6)$ in Fig. 3e], $H^s$ has two chiral symmetry operators $S^{sx}$ and $S^{sy}$ because Eq. (13) now applies to both the x and y directions. Again, using Eq. (15), we find that $N_{\Gamma,X}^{sx}(2\pi/3, \pi/2) = 0$ and $N_{\Gamma,X}^{sy}(2\pi/3, \pi/2) = 8$. We confirm the protection by $S^{sy}$ based on the gapping of the double Dirac points with two on-site potential perturbations $\Delta\lambda_{m,n}^{s1} = (-1)^{m+n}\Delta$ and $\Delta\lambda_{m,n}^{s2} = (-1)^{m+(n-1)q_\phi}\Delta$, which both preserve $S^{sx}$ and spatial symmetries but break $S^{sy}$ for $H^s(\pi/2, \pi/3)$ (Fig. 3e). We note that inversion and chiral symmetries enable the appearance of double Dirac points (8-fold degenerate) in this study, different from the previously proposed double Dirac semimetals enforced by non-symmorphic symmetries[71, 72].

Dirac points also appear at the off-TRIM points inside the MBZ of $H^s$. In Fig. 3f with $H^s(2\pi/3, \pi/3)$, two Dirac points appear at $\boldsymbol{k} = (\pi/2, \pi/2)$ and $\boldsymbol{k} = (-\pi/2, -\pi/2)$. They transform pairwise into each other in four ways, i.e., time-reversal, particle-hole, inversion, and joint (magnetic translation $k_x \to k_x + 2\pi/q_\phi$ with $C_2^x = IM_x$ rotation) operations. We test the stability of the off-TRIM Dirac points with various perturbations. $\Delta\lambda_{m,n}^{s1}$, which breaks $I$ but respects the other internal and reflection symmetries for $H^s(2\pi/3, \pi/3)$, gaps the Dirac degeneracy. For another perturbation $\Delta\lambda_{m,n}^{s3} = (-1)^{m<=q/2}\Delta$ that breaks $I$ and $M_x$ but preserves their product $C_2^x$, as well as all internal symmetries, the two Dirac points split into four Weyl nodes along the $k_x = \pi/6$ line. In contrast, for $\Delta\lambda_{m,n}^{s2}$, which respects all symmetries, the Dirac points are locally stable at momenta $\boldsymbol{k} = (\pi/6, \pm k_y)$.

It is interesting to investigate the conditions under which Dirac points disappear from the MBZ edge, as in Fig. 3d and f. We conjecture that the conditions are

$$\frac{q_{\phi(\theta)}}{q_{\theta(\phi)}} = 2 \quad \text{and} \quad q_{\theta(\phi)} \text{ is odd} \tag{16}$$

for both $H^L$ and $H^s$. Eq. (16) also guarantees the presence of chiral symmetry. This conjecture, which still calls for a rigorous proof, has supporting evidence from numerical tests, e.g., compare Fig. 3c with d and Fig. 3e with f. One special case of Eq. (16) worth mentioning is when $\theta + \phi = \pi$ (Fig. 3d and f). In this case, Im $D^L$ (Fig. 3d3), Re $D^s$ (Fig. 3f2), and Im $D^s$ (Fig. 3f3) permit simple analytical expressions. For example, in Fig. 3f2 and f3, Re $D^s \propto \sin^2 q_\theta k_x - \sin^2 q_\theta k_y$ and Im $D^s \propto -\cos q_\theta k_x \cos q_\theta k_y$, respectively.

The symmetric-gauge model $H^s$ becomes a square lattice if $\theta = \phi$. This special case enables richer degeneracies, which we briefly enumerate below. For $(\theta, \phi)/2\pi = (1/4, 1/4)$, a double Dirac point with 8-fold degeneracy appears at the $\Gamma$ point, which is again protected by chiral and inversion symmetries with $N_\Gamma^s(\pi/2, \pi/2) = 8$ [cf. Eq. (15)]. For $(\theta, \phi)/2\pi = (1/6, 1/6)$, a bulk Dirac nodal line connects multiple $\Gamma$ points along the $\Gamma - M$ direction in the MBZ. The nodal line, 4-fold degenerate at general momenta, becomes a triple Dirac point with 12-fold degeneracy at $\Gamma$ and a double Dirac point at $M$. For $(\theta, \phi)/2\pi = (1/8, 1/8)$ or $(3/8, 3/8)$, a quadruple Dirac point with 16-fold degeneracy appears at $\Gamma$. We are not aware of the construction of triple and quadruple Dirac points in the literature. We will investigate these aspects in future works.

After examining the degeneracies at half-filling, we discuss the gapped phases of the models. As an example, we choose $(\theta, \phi)/2\pi = (1/3, 1/6)$ for $H^L$ and $H^s$ and calculate their edge spectra (Fig. 4) by truncating one direction while maintaining periodicity in the other direction. Multiple bulk gaps emerge and can be categorized as the quantum spin Hall (QSH, $\mathbb{Z}_2$-odd) and normal band insulating (NI, $\mathbb{Z}_2$-even; we do not differentiate trivial and obstructed atomic insulators here) phases, as evident from the counting of the intersections between the edge modes and a given in-gap Fermi level within half of the MBZ. Recalling the bulk-edge correspondence, we confirm the QSH phases by calculating the 1D Wannier spectra (Fig. 4a2, b3, and b4)[73] of the bulk, which exhibit the typical winding of the Berry phases.

## Discussion

We have introduced and studied two distinct non-Abelian generalizations of the Hofstadter model in the Landau and symmetric gauges. We have analytically obtained the genuine non-Abelian condition, $[\Theta\Phi, \Phi\Theta] \neq 0$, shared by the two non-Abelian models by examining the commutativity of their arbitrary loop operators. This condition also dictates whether the bulk spectra reduce to the original Hofstadter butterfly. Protected by internal and spatial symmetries, a wide variety of Weyl and Dirac (single, double, triple, and quadruple) degeneracies occur at zero energy. Under nearest-neighbour coupling only, these models realize $\mathbb{Z}_2$ topological insulators in their bulk gaps.

Theoretically, the non-Abelian generalizations of the Hofstadter model can be extended to three dimensions (e.g., see[74]). We introduce a gauge potential $((l \pm n)\theta\sigma_x, (m \pm l)\phi\sigma_y, (n \pm m)\psi\sigma_z)$ in a cubic lattice $(m, n, l)$. This gauge potential gives rise to a Hamiltonian

$$H^{3D}(\theta, \phi, \psi) = -\sum_{m,n,l} t_x c^\dagger_{m+1,n,l} e^{i(l\pm n)\theta\sigma_x} c_{m,n,l} + t_y c^\dagger_{m,n+1,l} e^{i(m\pm l)\phi\sigma_y} c_{m,n,l} \quad (17)$$
$$+ t_z c^\dagger_{m,n,l+1} e^{i(n\pm m)\psi\sigma_z} c_{m,n,l} + \text{H.c.}$$

This three-dimensional system describes three pairs of Hofstadter butterflies that are spin-orbit coupled. In fact, every 2D surface (*xy*, *xz*, or *yz*) corresponds to the 2D non-Abelian model $H^s$. Aside from the first-order topological phases in two and three dimensions, it would be interesting to consider how these non-Abelian ingredients can be utilized to construct high-order topological phases[75] therein.

There are several directions for experimental demonstration of these models with photonics. First, established platforms, such as coupled laser-written waveguide arrays and integrated coupled resonators, could enable direct real-space realization (e.g., the possible scheme in Fig. S2). This requires an expansion of the internal degrees of freedom, such as exploiting a mode degeneracy, in these optical systems. Synthetic gauge fields that break time-reversal symmetry could also be created therein, such as by using integrated phase modulators and polarization/mode converters with lithium niobates and/or magnetic-optic materials. In addition, to achieve some time-reversal-invariant phases, it could be sufficient to engineer reciprocal phases, such as through a propagating phase delay[13] in different bases. Second, it is possible to realize the models in synthetic dimensions[76]. A synthetic dimension, such as frequency, can play the role of lattice sites and maintain the internal degrees of freedom. The dimensionality of the models can be encoded by driving the system with multiple frequencies. Alternatively, both non-Abelian models could be studied in a pumping experiment. Analogous to the Abelian Hofstadter model, $H^L$ and $H^S$ (after a gauge transformation) allow the replacement of one of the momenta with pumping[77]. These dimension-reduced approaches could help realize the models with relative ease in future experiments.

## Materials and methods

### A: Detailed Discussions on Genuine non-Abelian conditions

In this section, we prove the genuine non-Abelian condition Eq. (8) for both $H^L$ and $H^S$.

In our proposed models $H^L$ and $H^S$, the commutativity of arbitrary real-space Berry curvatures is equivalent to that of arbitrary loop operators of unit plaquettes, except for some possible trivial counterexamples. Although it is straightforward that for matrices $A$ and $B$, $AB = BA$ implies $e^A e^B = e^B e^A$, the converse is not always true. It has been proven[78] that for bounded operators $A$ and $B$ on a Banach space with $2\pi i$-congruence-free spectra, $e^A e^B = e^B e^A$ if and only if $AB = BA$. Here, $2\pi i$-congruence-free spectra refer to a set of eigenvalues $\Lambda$ where there are no two different elements $\{\lambda_1, \lambda_2\} \in \Lambda$ such that $\lambda_1 = \lambda_2 \bmod 2\pi i$. For our models $H^L$ and $H^S$, their loop operators are always SU(2) rotations. Therefore, the eigenvalues of real-space Berry curvatures are always angles $\pm\gamma$. As a result, $\pm\gamma$ are guaranteed to be 0 or $\pm\pi$ if they are $2\pi$-congruent. In both cases, the associated loop operator is the identity matrix. Therefore, for the equivalence between the commutativity of loop operators and Berry curvatures in our models, the counterexamples (i.e., commutative loop operators but non-commutative Berry curvatures) are trivial because there is no net rotation. These trivial counterexamples are a result of the non-uniqueness of the matrix logarithm (e.g., $e^{i\pi\sigma_{0,x,y,z}} = -\sigma_0$).

After ruling out the possible trivial counterexamples, we next examine the commutativity of arbitrary loop operators of unit plaquettes to derive the genuine non-Abelian conditions for our lattice models. We reiterate our compact notations introduced in the main text. We define $\Theta_m \equiv \exp(im\theta\sigma_y)$ and $\Phi_m \equiv \exp(im\phi\sigma_z)$ for the link variables. We also set $\Theta \equiv \Theta_1$ and $\Phi \equiv \Phi_1$. We first introduce a useful lemma.

**Lemma A.1** $\forall m, n \in \mathbb{Z}, [\Theta_m \Phi_m, \Theta_n \Phi_n] = 0$ *if and only if* $[\Theta\Phi, \Phi\Theta] = 0$.

*Proof:*

*Necessity*: $[\Theta_m \Phi_m, \Theta_n \Phi_n] = 0 \Rightarrow [\Theta_1 \Phi_1, \Phi_1 \Theta_1] = 0$

By choosing $m = 1$, $n = -1$ $[\Theta_1 \Phi_1, \Theta_{-1} \Phi_{-1}] = 0$. For unitary matrices $A$ and $B$, $[A, B] = 0 \leftrightarrow [A, B^\dagger] = 0$ because

$$AB^\dagger = B^\dagger BAB^\dagger = B^\dagger ABB^\dagger = B^\dagger A \qquad (18)$$

Therefore, $[\Theta_1 \Phi_1, \Phi_1 \Theta_1] = 0$. In addition, it can be readily shown from Eq. (18) that

$$[\Theta_m\Phi_m, \Theta_n\Phi_n] = 0 \leftrightarrow \Theta_m\Phi_{m+n}\Theta_n = \Phi_n\Theta_{m+n}\Phi_m \tag{19}$$

which will also be used frequently in the following.

*Sufficiency:* $[\Theta_1\Phi_1, \Phi_1\Theta_1] = 0 \Rightarrow [\Theta_m\Phi_m, \Theta_n\Phi_n] = 0$

Use mathematical induction. We know that $[\Theta_1\Phi_1, \Theta_{-1}\Phi_{-1}] = 0$, $[\Theta_1\Phi_1, \Theta_0\Phi_0] = 0$, and $[\Theta_{-1}\Phi_{-1}, \Theta_0\Phi_0] = 0$. We next prove that if $[\Theta_m\Phi_m, \Theta_n\Phi_n] = 0$ holds for any $(m', n')$ that satisfies $|m'| \leq m$ and $|n'| \leq n$, then $[\Theta_{m+1}\Phi_{m+1}, \Theta_n\Phi_n] = 0$ and $[\Theta_m\Phi_m, \Theta_{n+1}\Phi_{n+1}] = 0$ also hold.

$$\begin{aligned}
&[\Theta_{m+1}\Phi_{m+1}, \Theta_n\Phi_n] \\
&= [\underline{\Theta_1\Theta_m\Phi_m\Phi_1}, \Theta_n\Phi_n] \\
&= [\Theta_1, \Theta_n\Phi_n]\Theta_m\Phi_{m+1} + \Theta_{m+1}\Phi_m[\Phi_1, \Theta_n\Phi_n] \\
&= \Theta_n[\Theta_1, \Phi_n]\Theta_m\Phi_{m+1} + \Theta_{m+1}\Phi_m[\Phi_1, \Theta_n]\Phi_n
\end{aligned} \tag{20}$$

We have used the condition that $[\Theta_m\Phi_m, \Theta_n\Phi_n] = 0$. Now, we need to prove that $\Theta_n[\Theta_1, \Phi_n]\Theta_m\Phi_{m+1} = \Theta_{m+1}\Phi_m[\Theta_n, \Phi_1]\Phi_n$, i.e., $[\Theta_1, \Phi_n]\Theta_m\Phi_m\Phi_{1-n} = \Theta_{1-n}\Theta_m\Phi_m[\Theta_n, \Phi_1]$.

$$\begin{aligned}
\text{LHS} &= (\Theta_1\Phi_n - \Phi_n\Theta_1)\Theta_m\Phi_m\Phi_{1-n} \\
&= \Theta_{1-n}(\Theta_n\Phi_n - \Theta_{n-1}\Phi_n\Theta_1)\Theta_m\Phi_m\Phi_{1-n} \\
&= \Theta_{1-n}(\underline{\Theta_n\Phi_n\Theta_m\Phi_m}\Phi_{1-n} - \underline{\Theta_{n-1}\Phi_n\Theta_1}\Theta_m\Phi_m\Phi_{1-n}) \\
&\hspace{5cm} {\scriptstyle [\Theta_m\Phi_m, \Theta_n\Phi_n] = 0, \ [\Theta_1\Phi_1, \Theta_n\Phi_n] = 0} \\
&= \Theta_{1-n}(\underline{\Theta_m\Phi_m\Theta_n\Phi_n}\Phi_{1-n} - \Phi_1\Theta_n\Phi_{n-1}\Theta_m\Phi_m\Phi_{1-n}) \\
&= \Theta_{1-n}(\Theta_m\Phi_m\Theta_n\Phi_1 - \Phi_1\Theta_n\underline{\Phi_{n-1}\Theta_m\Phi_{m+1-n}}) \\
&\hspace{5cm} {\scriptstyle [\Theta_m\Phi_m, \Theta_{n-1}\Phi_{n-1}] = 0} \\
&= \Theta_{1-n}(\Theta_m\Phi_m\Theta_n\Phi_1 - \Phi_1\Theta_n\underline{\Theta_{m+1-n}\Phi_m\Theta_{n-1}}) \tag{21} \\
&= \Theta_{1-n}(\Theta_m\Phi_m\Theta_n\Phi_1 - \Phi_1\Theta_{m+1}\Phi_m\Theta_{n-1}) \\
&= \Theta_{1-n}(\Theta_m\Phi_m\Theta_n\Phi_1 - \underline{\Phi_1\Theta_{m+1}\Phi_m}\Theta_{n-1}) \\
&\hspace{5cm} {\scriptstyle [\Theta_m\Phi_m, \Theta_1\Phi_1] = 0} \\
&= \Theta_{1-n}(\Theta_m\Phi_m\Theta_n\Phi_1 - \underline{\Theta_m\Phi_{m+1}\Theta_1}\Theta_{n-1}) \\
&= \Theta_{1-n}(\Theta_m\Phi_m\Theta_n\Phi_1 - \Theta_m\Phi_m\Phi_1\Theta_n) \\
&= \Theta_{1-n}\Theta_m\Phi_m(\Theta_n\Phi_1 - \Phi_1\Theta_n) \\
&= \Theta_{1-n}\Theta_m\Phi_m[\Theta_n, \Phi_1] = \text{RHS}
\end{aligned}$$

∎

**Theorem A.2** *The non-Abelian Hofstadter model $H^L$ becomes genuinely Abelian if and only if $[\Theta\Phi, \Phi\Theta] = 0$.*

*Proof:*

The unit plaquette loop operators in the Landau-gauge model are

$$W^{\text{L}}_{m,n} = \Phi_{-m}\Theta\Phi_{m+1} \tag{22}$$

*Necessity*: We can simply choose two unit plaquettes $W^{\text{L}}_{0,n} = \Theta\Phi$ and $W^{\text{L}}_{-1,n} = \Phi\Theta$. Therefore, $[W^{\text{L}}_{m,n}, W^{\text{L}}_{m',n'}] = 0 \Rightarrow [\Theta\Phi, \Phi\Theta] = 0$.

*Sufficiency*: The loop operator can be rewritten as

$$W^{\text{L}}_{m,n} = \Phi_{-m}\Theta_{-m} \cdot \Theta_{m+1}\Phi_{m+1} \tag{23}$$

Recalling Lemma A.1 and Eq. (18), it is evident that $[W^{\text{L}}_{m,n}, W^{\text{L}}_{m',n'}] = 0$.

∎

**Theorem A.3** *The non-Abelian Hofstadter model $H^{\text{s}}$ becomes genuinely Abelian if and only if $[\Theta\Phi, \Phi\Theta] = 0$.*

*Proof:*

The loop operator for the symmetric-gauge model is

$$W^{\text{s}}_{m,n} = \Phi_{-m}\Theta_{n+1}\Phi_{m+1}\Theta_{-n} \tag{24}$$

*Necessity*: We can simply choose unit plaquettes $W^{\text{s}}_{m,-1} = W^{\text{s}}_{-1,n} = \Phi\Theta$. Therefore, $[W^{\text{s}}_{m,n}, W^{\text{s}}_{m',n'}] = 0 \Rightarrow [\Theta\Phi, \Phi\Theta] = 0$.

*Sufficiency*: We rewrite the loop operator Eq. (24) as

$$W^{\text{s}}_{m,n} = \Phi_{-m}\Theta_{-m} \cdot \Theta_{m+n+1}\Phi_{m+n+1} \cdot \Phi_{-n}\Theta_{-n} \tag{25}$$

Recalling Lemma A.1 and Eq. (18), it is evident that $[W^{\text{s}}_{m,n}, W^{\text{s}}_{m',n'}] = 0$.

∎

## B: Genuine non-Abelian conditions and the re-emergence of the Hofstadter butterfly

We next establish the equivalence between the spectral re-emergence of the Hofstadter butterfly and the genuine Abelian condition [cf. Eq. (8)].

For this, we first consider the simplest case when $\theta = 0$ for both $H^{\text{L}}$ and $H^{\text{s}}$. There is a permutation matrix $P$ that can block diagonalize the Hamiltonian as $PHP^{\dagger} = H^{\uparrow}_0(\phi) \oplus H^{\downarrow}_0(-\phi)$, where $H^{\uparrow,\downarrow}$ are the Abelian Hofstadter Hamiltonians [Eq. (1)] for up and down spins, respectively. When $\phi = 0$, the same argument applies after a basis change of $\sigma_y \to \sigma_z$.

### Landau-gauge model $H^{\text{L}}$

We have shown in the main text that the necessary and sufficient condition for a pair of decoupled butterflies is that there exists a global unitary transformation $U$ that can simultaneously diagonalize all $2 \times 2$ block diagonal terms $H^{\text{L}}_{m,m}$ of the Hamiltonian. This is equivalent to the requirement that all $H^{\text{L}}_{m,m}$ commute, i.e., the commutator

$$[e^{ik_y}\Theta_m\Phi_m + \text{H.c.}, e^{ik_y}\Theta_n\Phi_n + \text{H.c.}] = 0 \tag{26}$$

for an arbitrary choice of $m$, $n$, and $k_y$. Eq. (26) can be rearranged as

$$[e^{ik_y}\Theta_m\Phi_m, e^{ik_y}\Theta_n\Phi_n] + [e^{ik_y}\Theta_m\Phi_m, e^{-ik_y}\Phi_{-n}\Theta_{-n}] - \text{H.c.} = 0$$
$$e^{2ik_y}[\Theta_m\Phi_m, \Theta_n\Phi_n] + [\Theta_m\Phi_m, \Phi_{-n}\Theta_{-n}] - \text{H.c.} = 0 \quad (27)$$

Therefore, both $e^{2ik_y}[\Theta_m\Phi_m, \Theta_n\Phi_n]$ and $[\Theta_m\Phi_m, \Phi_{-n}\Theta_{-n}]$ must be real and symmetric. Eq. (18) shows that $[\Theta_m\Phi_m, \Theta_n\Phi_n]$ and $[\Theta_m\Phi_m, \Phi_{-n}\Theta_{-n}]$ are equivalent. Since Eq. (26) applies to an arbitrary $k_y$, we must have $[\Theta_m\Phi_m, \Theta_n\Phi_n] = 0$.

As such, $H^L$ reduces to a direct sum of two Abelian Hofstadter Hamiltonians if and only if all links commute, i.e., $[\Theta_m\Phi_m, \Theta_n\Phi_n] = 0$, which is equivalent to the genuine non-Abelian condition $[\Theta\Phi, \Phi\Theta] = 0$ (see Lemma A.1).

**Symmetric-gauge model $H^S$**

A local gauge transformation $U_{m,n} = \Theta_{-mn}$ can transform the symmetric-gauge model $H^S$ into the Landau gauge, where the complex link variables in the $x$ direction become real hopping. Now, the transformed link variable in the $y$ direction between $(m, n)$ and $(m, n+1)$ is

$$L^S_{m,n} = \Theta_{-mn}\Phi_m\Theta_{m(n+1)}$$
$$= \Theta_{-mn}\Phi_{-mn} \cdot \Phi_{m(n+1)}\Theta_{m(n+1)} \quad (28)$$

Again, we drop the notation for the non-trivial hopping $y$ direction, and we emphasize that $L^S_{m,n}$ and $L^L_{m,n} = \Theta_m\Phi_m$ are different. After the transformation, the Hamiltonian can be written in a basis where the complex non-Abelian hoppings all appear in the $2 \times 2$ block diagonal matrices. We first examine the commutativity of the transformed link variables [Eq. (28)] and introduce a theorem.

**Theorem B.1** $[L^S_{m,n}, L^S_{m',n'}] = 0$ *if and only if* $[\Theta\Phi, \Phi\Theta] = 0$.

*Proof:*

Necessity: We have

$$L^S_{1,-1} = \Theta\Phi \quad (29)$$
$$L^S_{1,-1} = \Theta_{-1}\Phi_{-1} \quad (30)$$

Therefore,

$$[L^S_{m,n}, L^S_{m',n'}] = 0 \rightarrow [\Theta\Phi, \Theta_{-1}\Phi_{-1}] = 0 \leftrightarrow [\Theta\Phi, \Phi\Theta] = 0 \quad (31)$$

The sufficiency is also evident based on Lemma A.1 and Eqs. (18) and (28).

∎

Using the same argument as in the previous section for the Landau-gauge model $H^L$, the spectral re-emergence of the Hofstadter butterfly is equivalent to the condition that all the link variables [Eq. (28)] are commutative. Theorem B.1 shows that such a condition is also equivalent to the genuine Abelian condition $[\Theta\Phi, \Phi\Theta] = 0$.

**C: Chiral symmetry**

**Landau-gauge model $H^L$**

We next show that the Landau-gauge model has chiral symmetry when $q_\theta q_\phi$ is even and $p_\theta \tilde{q}_\phi + p_\phi \tilde{q}_\theta$ is odd, where $\tilde{q}_{\theta(\phi)} \equiv q_{\theta(\phi)}/\gcd(q_\theta, q_\phi)$ and gcd is the greatest common divisor.

Rather than discussing $E(k_x, k_y) \to -E(k_x, k_y)$ directly, we examine the chiral symmetry via the conditions for $E(k_x, k_y) \to -E(k_x + \pi, k_y + \pi)$ and $E(k_x, k_y) \to E(k_x + \pi, k_y + \pi)$. The Landau-gauge model has a Harper equation given by

$$E\Psi_{m,n} = -t_x(\Psi_{m+1,n} + \Psi_{m-1,n}) - t_y(\Theta_m \Phi_m \Psi_{m,n+1} + \Phi_{-m}\Theta_{-m}\Psi_{m,n-1}) \tag{32}$$

Under Bloch's theorem, $\Psi_{m,n} = e^{ik_x m} e^{ik_y n} u_m$ and $u_m = u_{m+q}$, where $q = \mathrm{lcm}(q_\theta, q_\phi)$. Eq. (32) becomes

$$Eu_m = -t[u_{m+1}e^{ik_x} + u_{m-1}e^{-ik_x} + (e^{ik_y}\Theta_m \Phi_m + \mathrm{H.c.})u_m] \tag{33}$$

We adopt a trial wavefunction $\tilde{\Psi}_{m,n} = e^{i(k_x+\pi)m} e^{i(k_y+\pi)n} u_m$ and obtain

$$-Eu_m = -t\left\{u_{m+1}e^{i(k_x+\pi)} + u_{m-1}e^{-i(k_x+\pi)} + \left[e^{i(k_y+\pi)}\Theta_m \Phi_m + \mathrm{H.c.}\right]u_m\right\} \tag{34}$$

which is exactly equivalent to Eq. (32). Therefore, we have symmetry $E(k_x, k_y) \to -E(k_x + \pi, k_y + \pi)$ regardless of the choice of $\theta$ and $\phi$. This property is the same as that of the Abelian model $H^0$.

Next, we discuss the conditions for $E(k_x, k_y) \to E(k_x + \pi, k_y + \pi)$. In the $k_x$ direction, $k_x$ and $k_x + 2\pi/q$ are equivalent. If $q$ is even (i.e., at least one of $q_\theta$ and $q_\phi$ is even), $k_x$ and $k_x + \pi$ become equivalent, and therefore, $E(k_x, k_y) \to E(k_x + \pi, k_y)$.

For $k_y$, we consider the lattice site $q/2 + 1$. The associated links are $e^{i(q/2+1)\theta \sigma_y} e^{i(q/2+1)\phi \sigma_z} = e^{i\theta \sigma_y} e^{i\phi \sigma_z} e^{i\pi(p_\theta \tilde{q}_\phi + p_\phi \tilde{q}_\theta)}$, where $\tilde{q}_{\theta(\phi)} \equiv q_{\theta(\phi)}/\gcd(q_\theta, q_\phi)$ and gcd is the greatest common divisor. We also use the fact that $e^{i\pi\sigma_z} = e^{i\pi\sigma_y} = e^{i\pi}\sigma_0$. Therefore, if $q_\theta q_\phi$ is even and $p_\theta \tilde{q}_\phi + p_\phi \tilde{q}_\theta$ is odd, then the link at site $q/2 + 1$ becomes $e^{i\pi} e^{i\theta \sigma_y} e^{i\phi \sigma_z}$. This implies that $E(k_x, k_y) \to E(k_x, k_y + \pi)$ by the transformation $u_m \to u_{m+q/2}$.

Taking together the conditions for $E(k_x, k_y) \to -E(k_x + \pi, k_y + \pi)$ and $E(k_x, k_y) \to E(k_x + \pi, k_y + \pi)$, the condition for $H^L$ to have chiral symmetry is that $q_\theta q_\phi$ is even and $p_\theta \tilde{q}_\phi + p_\phi \tilde{q}_\theta$ is odd. For example, when $q_\theta = q_\phi$ (and they are both even), $p_\theta \tilde{q}_\phi + p_\phi \tilde{q}_\theta = p_\theta + p_\phi$, which is guaranteed to be even (recall the incommensurability requirement). In another example, $(\theta, \phi)/2\pi = (1/4, 1/12)$, where $p_\theta \tilde{q}_\phi + p_\phi \tilde{q}_\theta = 4$ is again even. In both examples, $H^L$ does not have chiral symmetry.

**Symmetric-gauge model $H^S$**

Next, we show that the symmetric-gauge model $H^S$ has chiral symmetry when $q_\theta q_\phi$ is even. This requirement is weaker than that of the Landau-gauge model $H^L$. Again, we examine the chiral symmetry via the conditions for $E(k_x, k_y) \to -E(k_x + \pi, k_y + \pi)$ and $E(k_x, k_y) \to E(k_x + \pi, k_y + \pi)$. The Harper equation for $H^s$ is

$$E\Psi_{m,n} = -t_x(\Theta_{-n}\Psi_{m+1,n} + \Theta_n\Psi_{m-1,n}) - t_y(\Phi_m\Psi_{m,n+1} + \Phi_{-m}\Psi_{m,n-1}) \tag{35}$$

Under Bloch's theorem, $\Psi_{m,n} = e^{ik_x m}e^{ik_y n}u_{m,n}$, $u_{m,n} = u_{m+q_\theta,n}$, and $u_{m,n} = u_{m,n+q_\phi}$. Thus, the magnetic Brillouin zone is $k_x \in [0, 2\pi/q_\phi)$ and $k_y \in [0, 2\pi/q_\theta)$. Then, Eq.(35) becomes

$$Eu_{m,n} = -t_x(e^{ik_x}\Theta_{-n}u_{m+1,n} + e^{-ik_x}\Theta_n u_{m-1,n})$$
$$-t_y(e^{ik_y}\Phi_m u_{m,n+1} + e^{-ik_y}\Phi_{-m}u_{m,n-1}) \tag{36}$$

We can examine the same transformed wavefunction $\tilde{\Psi}_{m,n} = e^{i(k_x+\pi)m}e^{i(k_y+\pi)n}u_{m,n}$, which also satisfies the Harper equation at energy $-E$. Therefore, $H^s$ also inherits the symmetry $E(k_x, k_y) \to -E(k_x + \pi, k_y + \pi)$, regardless of the choice of $q_\theta$ and $q_\phi$, from the Abelian Hofstadter model $H^0$.

We next show that $q_\theta q_\phi$ being even enables $E(k_x, k_y) \to E(k_x + \pi, k_y + \pi)$. Without loss of generality, we assume that $q_\theta$ (associated with the links in the $x$ direction) is even. Because $k_y \in [0, 2\pi/q_\theta)$, $k_y = 0$ and $k_y = \pi$ become equivalent, which renders $E(k_x, k_y) \to E(k_x, k_y + \pi)$. For $k_x$, since $q_\theta$ is even, the $x$-direction link at site $(m, q_\theta/2 + n)$ becomes equivalent to $\exp i(n\theta\sigma_y + \pi)$. Therefore, for an energy $E$ with eigenstate $\Psi_{m,n}(k_x, k_y)$, we have another state $\Psi_{m,n+q_\theta/2}(k_x + \pi, k_y)$. Taken together, the Hamiltonian $H^s$ has chiral symmetry when $q_\theta q_\phi$ is even.

### D: General properties of non-Abelian Hofstadter models

As we have noted in the main manuscript, a key distinction between our proposed models and other non-Abelian genearlizations[51, 58-62] is that our models describe two pairs of Hofstadter butterflies that are spin-orbit coupled. Below, we elaborate on a few other similarities and differences.

*Lattice configuration*: These non-Abelian Hofstadter models are all square lattice models with nearest-neighbour coupling and net Abelian and/or non-Abelian magnetic fluxes.

*Gapless zero modes:* The Abelian Hofstadter model $H^0$ is gapless at zero energy (see Fig. 3a-b). This property is also inherited by non-Abelian Hofstadter models (e.g., see Fig. 3c-f of our models and Ref.[58]).

*Gapped topological phases:* Depending on whether a non-Abelian Hofstadter model contains U(1) magnetic fluxes or not, the gapped topological phases can be categorized as quantum Hall[51, 58] or quantum spin Hall[59] phases, respectively. Our proposed models contain no U(1) links and therefore give rise to $\mathbb{Z}_2$ topological insulators.

## Acknowledgements

We thank Vincent Liu for collaborating on a related project. We thank Hrvoje Buljan, Liang Fu, and Ashvin Vishwanath for discussions. We thank Thomas Christensen, Vincent Liu, Hoi Chun Po, and Ziming Zhu for discussions and reading the manuscript. This work was supported by the Army Research Office under Cooperative Agreement W911NF-18-2-0048, NSF grant CCF-1640012, NSF grant DMR-1838412, AFRL contract FA8650-16-D-5403, the Air Force Office of Scientific Research under award numbers FA9550-20-1-0115 and FA9550-18-1-0133, the US Office of Naval Research (ONR) Multidisciplinary University Research Initiative (MURI) grant N00014-20-1-2325 on Robust Photonic Materials with High-Order Topological Protection, and the Charles E. Kaufman Foundation, a supporting organization of the Pittsburgh Foundation.

## Conflict of interests

The authors declare no conflict of interests.

## Contributions

Y.Y. performed the research and wrote the manuscript with input from all the authors.

**Supplementary information** accompanies the manuscript on the Light: Science & Applications website (http://www.nature.com/lsa)

**Figures with captions**

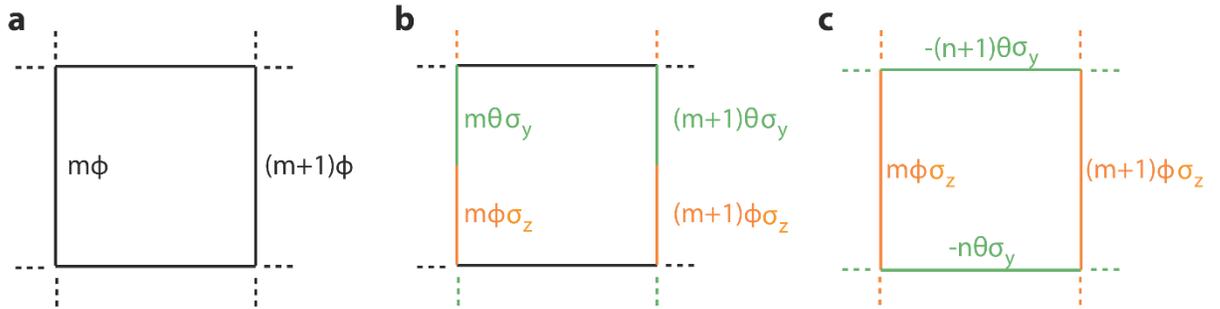

**Figure 1. Hofstadter model and its two non-Abelian generalizations. a.** Original Abelian Hofstadter model in the Landau gauge. **b.** Non-Abelian generalization in the Landau gauge [$H^L$ in Eq. (4)]. **c.** Non-Abelian generalization in the symmetric gauge [$H^s$ in Eq. (6)]. $H^L$ and $H^s$ are physically distinct models because of the non-commutativity of the gauge fields.

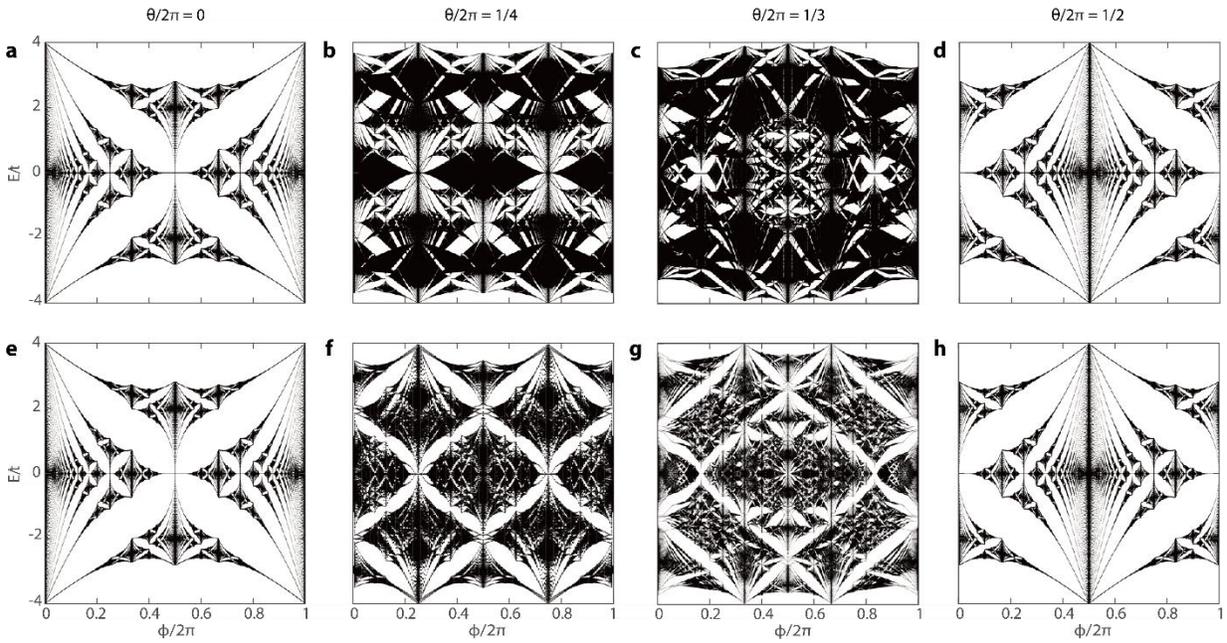

**Figure 2. Bulk spectra $E(\theta, \phi)$ of the Landau-gauge [Eq. (4); (a-d)] and symmetric-gauge [Eq. (6); (e-h)] non-Abelian models.** Examples for $\theta/2\pi \in \{0, 1/4, 1/3, 1/2\}$ are shown, and we choose $q_\phi = 1023$ to achieve high resolution. Eigenenergies for all momenta all overlayed at a given $(\theta, \phi)$. For $\theta/2\pi \in \{0, 1/2\}$, the models are Abelian, and the spectra reduce to two independent copies of the Hofstadter butterfly (**a**, **d**, **e**, and **h**). Among all cases, the spectra have a periodicity of $2\pi$ in the $\phi$ direction. The spectra are symmetric with respect to $E = 0$.

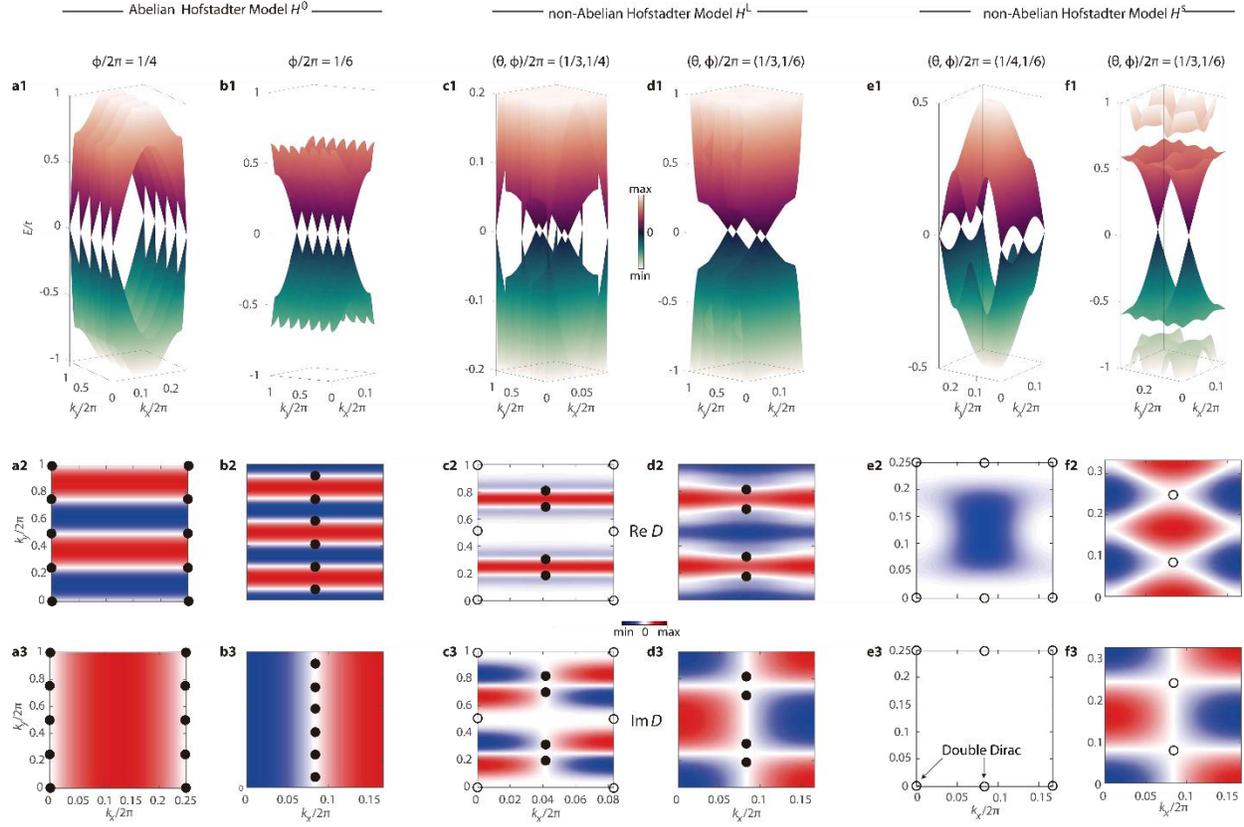

**Figure 3. Weyl and Dirac degeneracy at zero energy.** Energy spectra near zero energy (top) [full spectra shown in Fig. S1] and real (middle) and imaginary (bottom) parts of the determinant of the reduced Hamiltonian $D \equiv \det h$ for the Abelian Hofstadter model $H^0$ (**a-b**), the non-Abelian Landau-gauge model $H^L$ (**c-d**), and the non-Abelian symmetric-gauge model $H^s$ (**e-f**). A common zero in Re $D$ (middle) and Im $D$ (bottom) corresponds to a Weyl (solid circle) or Dirac (open circle) degeneracy.

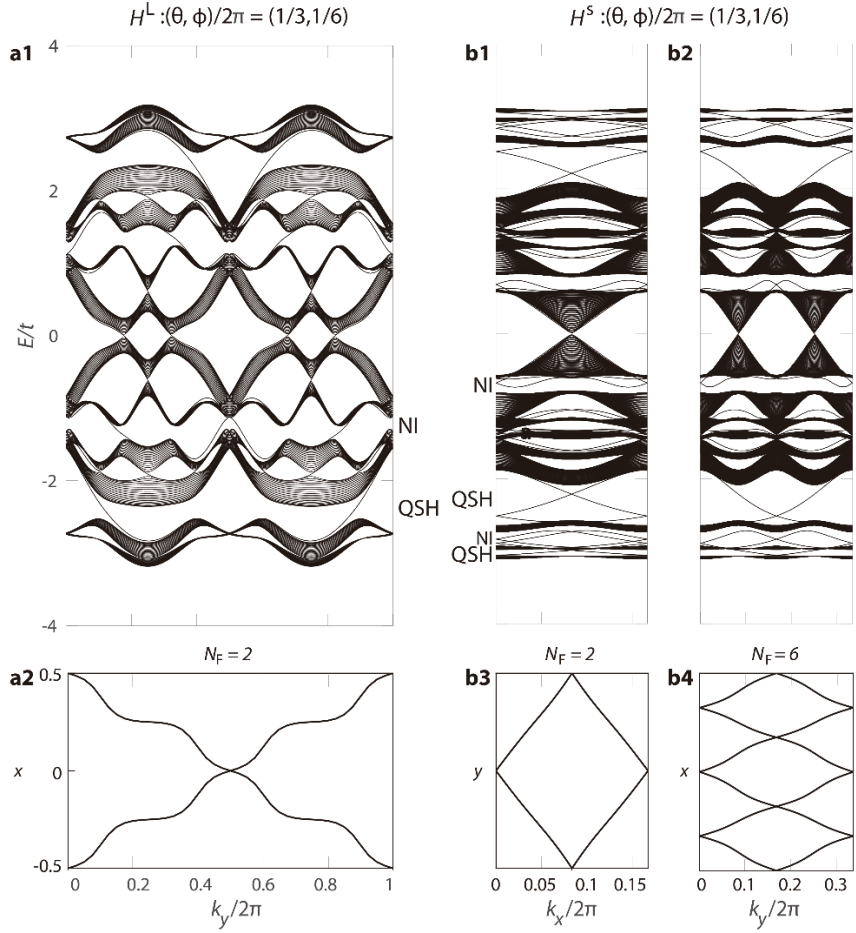

**Figure 4. Edge spectra and 1D Wannier spectra for the Landau-gauge (a) and symmetric-gauge models (b).** For $H^L$, we keep the periodicity in the $y$ direction (**4a1**). For $H^s$, we keep the periodicity in either the $x$ (**b1**) or $y$ (**b2**) direction. The topological phases associated with the bulk gaps are labelled as $\mathbb{Z}_2$-odd quantum spin Hall (QSH) insulators and normal band insulators (NI). We only label half of the gaps because chiral symmetry is present for $(\theta, \phi)/2\pi = (1/3, 1/6)$. The QSH phases are confirmed by the winding of the 1D Wannier spectra (**a2**, **b3**, and **b4**) for different gaps (labelled by the number of filled bands $N_F$).